\documentclass{article}

\usepackage[final]{neurips_2020}

\usepackage[utf8]{inputenc} 
\usepackage[T1]{fontenc}    
\usepackage{hyperref}       
\usepackage{url}            
\usepackage{booktabs}       
\usepackage{amsfonts}       
\usepackage{nicefrac}       
\usepackage{microtype}      
\usepackage{graphicx}

\usepackage{subfig}
\usepackage{amsmath}
\usepackage[title]{appendix}

\title{Lamina-specific neuronal properties promote robust, stable signal propagation in feedforward networks}

\author{
  Dongqi Han \\
  Cognitive Neurorobotics Research Unit\\
  Okinawa Institute of Science and Technology\\
  Okinawa, Japan \\
  \texttt{dongqi.han@oist.jp} \\
   \And
 Erik De Schutter \\
  Computational Neuroscience Unit\\
  Okinawa Institute of Science and Technology\\
  Okinawa, Japan \\
  \texttt{erik@oist.jp} \\
     \And
 Sungho Hong\thanks{Corresponding author} \\
  Computational Neuroscience Unit\\
  Okinawa Institute of Science and Technology\\
  Okinawa, Japan \\
  \texttt{shhong@oist.jp} \\
}

\begin{document}

\maketitle

\begin{abstract}
Feedforward networks (FFN) are ubiquitous structures in neural systems and have been studied to understand mechanisms of reliable signal and information transmission. In many FFNs, neurons in one layer have intrinsic properties that are distinct from those in their pre-/postsynaptic layers, but how this affects network-level information processing remains unexplored. Here we show that layer-to-layer heterogeneity arising from lamina-specific cellular properties facilitates signal and information transmission in FFNs. Specifically, we found that signal transformations, made by each layer of neurons on an input-driven spike signal, demodulate signal distortions introduced by preceding layers. This mechanism boosts information transfer carried by a propagating spike signal and thereby supports reliable spike signal and information transmission in a deep FFN. Our study suggests that distinct cell types in neural circuits, performing different computational functions, facilitate information processing on the whole.
\end{abstract}

\section{Introduction}
How different cell types in a neural system contribute to signal processing by the entire circuit is a prime question in neuroscience. Experimental investigations of this question are increasingly common, primarily due to advances in observing and manipulating neurons based on their genetic signature. Feedforward circuits are notable targets of those studies, since, in many systems, they have been observed to comprise cell groups or “layers” with properties distinct from those of other layers, in size, morphology, expression of membrane/intracellular mechanisms, etc. For example, in the \textit{Drosophila} antennal lobe (AL), projection neurons (PN) tend to show noisy firing, slow responses to the onset of olfactory receptor neuron (ORN) firing, and static voltage thresholds for spike generation whereas postsynaptic neurons of PNs in lateral horns (LHN) are less noisy, fire early, and have dynamical firing thresholds~\citep{Jeanne:2015di}. In the cerebellum, the granule cells are tiny neurons with simple morphology, but their postsynaptic targets, Purkinje cells, are large, with complex dendritic trees. Pre- and postsynaptic neurons having distinct intrinsic properties can be ubiquitously found in a wide variety of neural systems. These observations raise questions about the role of intrinsic properties and their laminar specificity. However, most theoretical and computational studies rarely take neuronal heterogeneity into account.

We addressed this question by studying the classical problem of how a spike signal, defined by the evoked firing of multiple neurons in one layer, can stably propagate through multiple downstream layers in an FFN~\citep{Diesmann:1999ina, vanRossum:2002vz, Reyes:2003gb, Kumar:2008ir, Kumar:2010dv, Kremkow:2010gx, Moldakarimov:2015gb, Joglekar:2018gq}. FFN is an important model for reliable information transfer between distant brain regions, since, if a need for new functional connections arises due to learning, long-distance axons for direct connections cannot grow in adult brains but can be added only through evolutionary processes. Instead, the connectome data suggest that brain regions tend to organize into clusters to allow for strong \textit{indirect} connectivity~\citep{oh2014mesoscale}, which can form FFNs. Stable propagation of spike signals in FFNs also plays a key role in models of conscious perception~\citep{Joglekar:2018gq, vanVugt:2018gk}, learning in deep artificial networks~\citep{schoengloz2016deep}, etc.

\begin{figure}[t]
    \centering
    \includegraphics[width=1.0\textwidth]{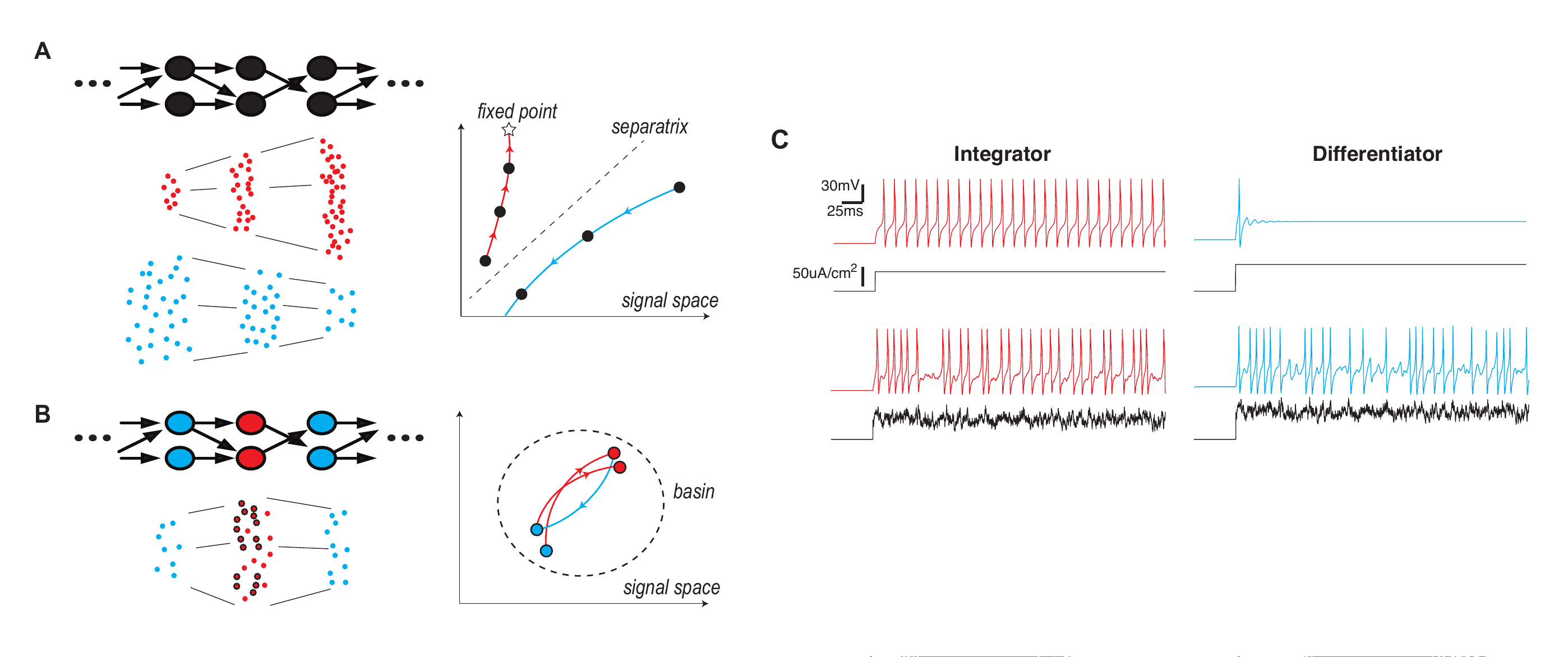}
    \caption{\textbf{Lamina-specific intrinsic properties stabilize information transmission in a neural network.} (A) Left: FFN with a single cell type (Top), and spikes at each layer, in two different modes of signal propagation (Bottom). One mode is amplification by progressively evoking more and more spikes (red dots), and the other is dissipation by gradually losing spikes (blue dots). Right: Trajectories of the two propagating signals in a signal space. The x- and y-axes represent independent signal characteristics, such as the number of spikes, temporal precision, etc. A star is a fixed point of neuronal signal transformation, and a dotted line is a separatrix separating the two modes. (B) Left: An FFN where neurons have lamina-specific intrinsic properties (Top). Each layer performs a layer-specific transformation, and can selectively transfer a subset of input spikes (circled red dots), ignoring those that cause signal distortion (Bottom). Right: Trajectory of a propagating signal in a signal space. The dotted circle surrounds a region (basin) where layer-specific transformations confine the propagating signal. (C) Behaviors of the two types of neurons used in this study with different dynamics of their spiking thresholds, shown by membrane potential response (color) to constant or fluctuating current injection (black).}
    \label{fig:1}
\end{figure}

Most of those studies assumed that FFNs have identical types of neurons, and thus each layer makes similar input/output transformations. In this case, an input-driven spike signal either gets stronger or weaker as it passes through layers, depending on the efficacy of output spike generation, given input spikes and also given the characteristics of the network input (Fig.~\ref{fig:1}(A), Left). Then, the signal eventually reaches a fixed point of layer-to-layer transformation or dissipates~\citep{Diesmann:1999ina, Reyes:2003gb, Moldakarimov:2015gb} (Fig.~\ref{fig:1}(A), Right). In this scenario, stable signal transmission is achieved by specific conditions for a non-trivial fixed point, which are often not robust for a wide range of initial signals. Also, irreversible signal distortion during propagation can cause the inevitable dissipation of information.

Introducing lamina-specific intrinsic properties in neurons can change this fundamentally (Fig.~\ref{fig:1}(B)). If each layer transforms a propagating signal in a different direction than the previous one, a fixed point will not exist in general. Instead, this prevents the repeated transformation of the signal in one direction, and the overall signal distortion over multiple layers can become smaller, compared to networks with identical layers. In particular, if the transformation by one layer is in the opposite or nearly opposite direction to those by its presynaptic layer, it can limit signal distortion across multiple layers and facilitate stable propagation (Fig.~\ref{fig:1}(B) Right). At the same time, information transfer also improves. Signal distortion at each layer will accumulate if neurons in each layer repeatedly encode similar preferred features of a network input, and this will cause irreversible loss of information. Contrarily, if the pre- and postsynaptic neurons encode distinct input features, more selective filtering on presynaptic output by postsynaptic neurons will demodulate distortions that presynaptic signal transformation introduced (Fig.~\ref{fig:1}(B) Left). In this manner, FFNs with heterogeneous, lamina-specific neuronal properties can show enhanced information transmission compared to homogeneous FFNs.

Here we demonstrate how robust, stable signal and information transmission arise from laminar specificity of cell-intrinsic properties by computational FFN models. We first introduce two types of neurons with different spiking dynamics, referred to as integrator and differentiator~\citep{Kumar:2010dv, Ratte:2013fg, Ratte:2014jr} (Fig.~\ref{fig:1}(C)). Using neural layers with such laminar specificity, we develop a model of the \textit{Drosophila} AL network with three layers of ORNs, PNs, and LHNs. We show that this model replicates key findings of a recent experimental study~\citep{Jeanne:2015di} that differences in spiking dynamics between PNs and LHNs can balance accuracy and speed in processing olfactory information, and furthermore that PN-to-LHN information transfer is nearly optimal. Then, we extend the model to a deep FFN and demonstrate robust and stable spike signal propagation, contrary to models with no laminar specificity in neuronal properties. Finally, we demonstrate that the speed of a propagating signal depends on the input signal property and, therefore, that deeper layers can use the latency coding for the input.

\section{Related work}
\label{chap:related_work}
Many computational studies have explored how a spike signal can reliably propagate in spiking neural networks. Most of them hypothesize that the FFNs consist of homogeneous excitatory neurons. Earlier work~\citep{Diesmann:1999ina, Kumar:2008ir, Kumar:2010dv, vanRossum:2002vz, Vogels:2005fc} investigated conditions of stable signal transmission in multi-layer networks with purely feedforward connections, mostly as a form of synchronized spikes. These conditions involve input signal properties~\citep{Diesmann:1999ina, Kumar:2008ir, Kumar:2010dv}, noise in neural dynamics~\citep{vanRossum:2002vz, Vogels:2005fc}, and connection strength and sparsity~\citep{Kumar:2008ir, Kumar:2010dv}. Recent studies also suggested that stable transmission can be enhanced by feedback connections~\citep{Moldakarimov:2015gb, Joglekar:2018gq}.

While heterogeneity in synaptic properties are often studied, heterogeneity of cellular properties in neural networks has been less emphasized. \cite{Kumar:2008ir} model neurons in a group with random distributions of intrinsic properties, and many network models \citep{yarden2017stimulus, Joglekar:2018gq} introduces heterogeneity of different areas so as to match physiological findings. However, these studies have not discussed how heterogeneous neuronal properties affect the entire network. Our work is original as we focus on the role of laminar-specific neuronal properties in FFNs, which have been rarely investigated in previous studies, despite their ubiquity in neural systems. Our paper is the first one, to our knowledge, to computationally investigate the role of laminar-specific neuronal properties for reliable signal transmission through FFNs.

\section{Methods}
We used conductance-based model neurons based on the Morris-Lecar mechanisms~\citep{Morris:1981iu, Prescott:2008bg}, given by
\begin{align}
C \frac{dV}{dt} &= - g_L (V - E_L) - g_K w (V - E_K) - g_{Na} m_\infty (V) (V-E_{Na}) + I_{stoch} + I_{input},\nonumber\\
\frac{dw}{dt} &= \phi_w \frac{w_\infty(V)-w}{\tau_w(V)}, \quad z_\infty(V) = \frac{1}{2} \left[ 1 + \tanh\left( \frac{V-\beta_z}{\gamma_z} \right) \right ] \quad (z = m, w),\nonumber\\
\tau_w(V) &= \cosh\left( \frac{V-\beta_w}{\gamma_w}\right)^{-1},
\label{eq:1}
\end{align}
where $V$ and $w$ are membrane potential and a gating variable for the $\mbox{K}^+$ channel. The first model, which we called the ``integrator'' neuron, had a high \textit{half-maximum voltage} of the $\mbox{K}^+$ channel, $\beta_w = 5$~mV. The other, ``differentiator'' neuron, had low $\beta_w=-19$~mV (See Section~\ref{chap:beta_w}). A special case is the ORN in the \textit{Drosophila} AL network model, which had $\beta_w=-23$~mV for stronger differentiator traits~\citep{nagel2011biophysical}. The other parameters, which are the same as in~\cite{Prescott:2008bg}, are shown in Appendix Table A1. Stochastic current $I_{stoch}$ represented noisy membrane potential fluctuation due to components that are absent in our model, such as background network inputs, an unknown noise source~\citep{Jeanne:2015di}, etc., and was given by an Ornstein-Uhlenbeck (OU) process, $d I_{stoch}/dt = -I_{stoch}/\tau_V + \sigma_V\xi$, where $\xi$ is a unit Gaussian noise, renewed each time step. $\tau_V=1$ ms, and $\sigma_V$ was tuned to experimental data of~\cite{Jeanne:2015di}.

The input current $I_{input}$ was either synaptic inputs or a common current injection to input layer neurons. Each synaptic input were conductance-based and modeled as a double exponential function: at each presynaptic spike at $t_s$, the synaptic current was
\begin{equation}
I_{input}(t) = g(t) (V-E_{syn}), \quad
g(t) = g_{syn} \left[ e^{-(t-t_s)/\tau_1} - e^{-(t-t_s)/\tau_2} \right] H(t-t_s),
\end{equation}
where $H(t)$ is a Heaviside function that $H(t) = 1$ if $t>0$ and $H(t) = 0$ otherwise. We used $\tau_1=0.5$ ms and $\tau_2=4$ ms, tuned $g_{syn}$ to experimental data as $\sigma_V$. All other parameters are in Appendix Table A1. In the simulation of the optogenetic experiments in~\citep{Jeanne:2015di}, the input layer neurons were injected
\begin{align}
    I_{input}=I_{amp} [e^{(-(t-t_0)/\tau_{act} )}-e^{(-(t-t_0)/\tau_{deact} )} ]H(t-t_0),
\end{align}
where $I_{amp}$=45 $\mu$A/c$\mbox{m}^2$, $\tau_{act}$=15 ms, and $\tau_{deact}$=50 ms. $t_0$=200 ms is a stimulus onset. See Appendix Table A2 for other parameters. We also performed a similar current injection to the input layer to generate dynamical spike signals (Fig.~\ref{fig:3}(D-E) and \ref{fig:4}(D)). Here we used the OU process,
\begin{align}
    dI_{input}/dt = (\mu_{input}-I_{input})/\tau_{input}+\sigma_e \xi, \label{eq:ou2}
\end{align}
where $\mu_{input}$=15 $\mu$A/c$\mbox{m}^2$, $\sigma_{input}$=7.5 $\mu$A/c$\mbox{m}^2$, and $\tau_{input}$=5 ms.

All the simulations were constructed and run in Python using the Brian2 simulator~\citep{Goodman:2008fb}.

\section{Results}

\subsection{Voltage-sensitive $\mbox{K}^+$ channels control dynamical input/output properties of neurons}
\label{chap:beta_w}
A recent experimental study showed that PNs and their feedforward targets, LHNs, have different intrinsic properties in \textit{Drosophila} AL network \citep{Jeanne:2015di}. We created the computational models of those neurons, described by Eq.~\eqref{eq:1}. Here a crucial parameter is $\beta_w$, the half-activation voltage of the $\mbox{K}^+$ channel. With lower $\beta_w$, the channel is more active at subthreshold voltages, shifting the balance between inward and outward currents around the spiking threshold. Then, the computational property of the neuron profoundly changes with strengthened differentiator-like traits, whereas higher $\beta_w$ promotes integrator-like behavior~\citep{Prescott:2008bg, Ratte:2013fg}. For example, typical repetitive firing with a sustained current input, seen in the integrator neurons with high $\beta_w$, was suppressed in those with the differentiators with low $\beta_w$. However, the differentiators showed robust sensitivity to the dynamic fluctuation in inputs, demonstrated by evoked firing (Fig.~\ref{fig:1}(C)). Crucially, our integrator and differentiator model neurons showed experimentally observed differences in the spiking thresholds. The spiking threshold was significantly more dynamic in the differentiators, just as in LHNs, whereas the integrators tended to have the static voltage threshold, as in PNs (see Appendix A1). These suggested that the integrator and differentiator neurons can be good models for \textit{Drosophila} PNs and LHNs, respectively.

\subsection{Lamina-specific neuronal properties are crucial for the \textit{Drosophila} AL network }

Using the model neurons, we developed a network model of the \textit{Drosophila} AL (see Appendix~\ref{ALDIFINT} for parameters). 40 ORNs projected to each PN, and nine PNs projected to each of nine LHNs~\citep{Jeanne:2015di}. Each layer contained 100 replicas of these, corresponding to 100 “trials” of an experiment, resulting in 4,000 ORNs, 900 PNs, and 900 LHNs in an entire network. We tuned the parameters, such as injected current, synaptic conductances and $\sigma_V$ for each layer, to match experimental measurements for i) the mean spontaneous firing rates in all layers and higher cell-to-cell variability in PN firing rates~\citep{Jeanne:2015di}, ii) mean peak firing rates, and iii) rate of decrease in mean LHN firing rate.

\begin{figure}
    \centering
    \includegraphics[width=1.0\textwidth]{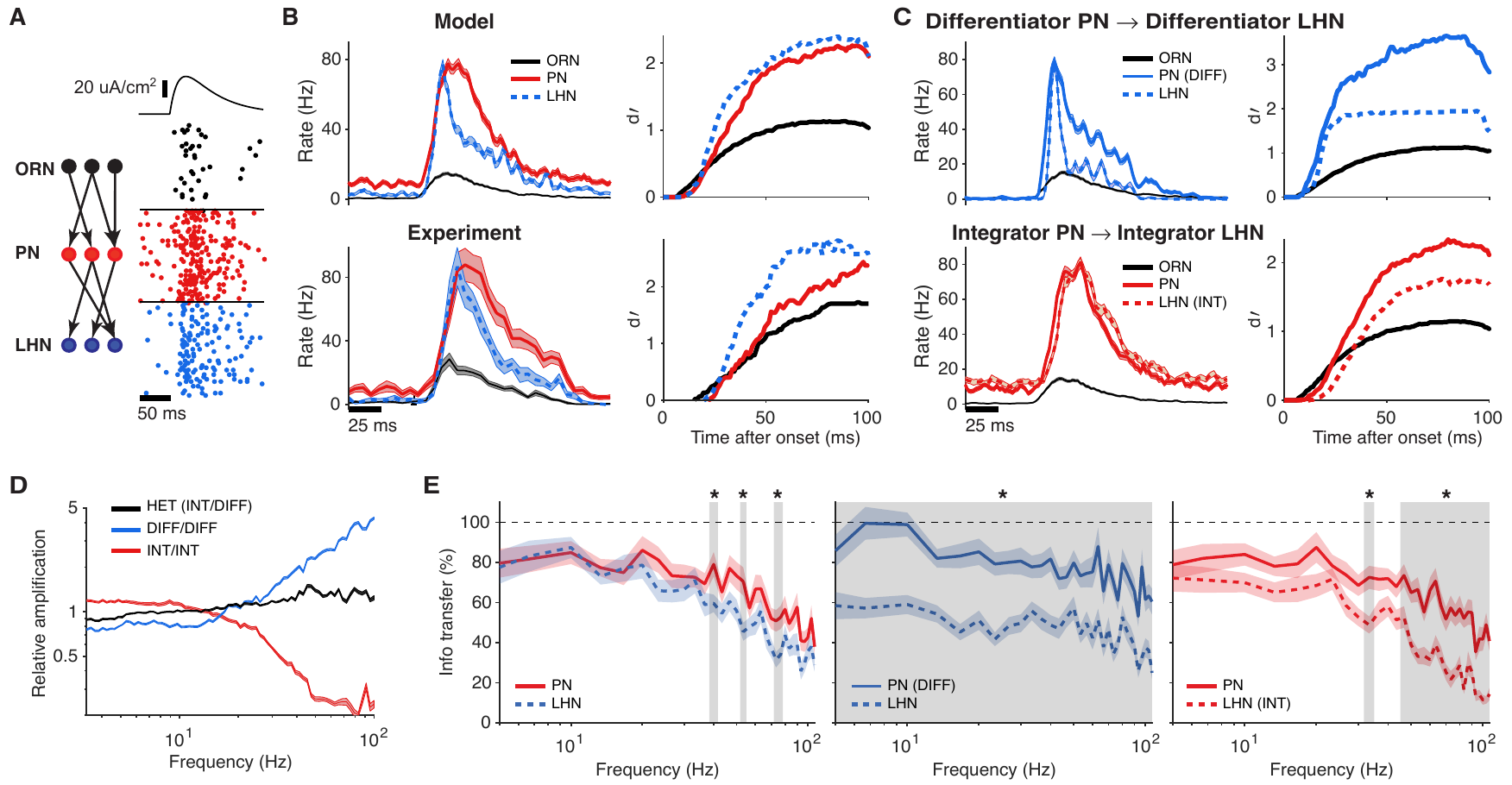}
    \caption{\textbf{Lamina-specific neuronal properties boost information transfer in the AL network.} (A) Schematic diagram of the network model (Left), and spikes from a simulation with a current input to ORNs on top (Right). 40 trials are shown for one example neuron in each layer. (B) Left: Average firing rates from the simulation (Top) and experimental data (Bottom). Right: ${d^\prime}$ for detecting an input to ORNs at each layer, computed from the same data as Left. (C) Same plots as B, with a model with differentiator PNs (Top), and with integrator LHNs (Bottom). (D) Spectral power amplification, $P_\text{LHN}(\omega)/P_\text{ORN}(\omega)$, normalized by total power. $P(\omega)$ is a power spectral density of mean firing rate. (E) Information transfer from ORNs to PNs (solid) and LHNs (dotted). Black dotted lines represent 100\% information transfer. Grey regions and stars represent frequency bands with significant differences between PNs and LHNs (*: \textit{P}<0.01, Student \textit{t}-test). Data are mean$\pm$SEM.}
    \label{fig:3}
\end{figure}

In the simulation of the optogenetic experiments in~\cite{Jeanne:2015di}, our network model reproduced the key features of experimental results. When ORNs were given a common current input that simulates optogenetic stimulation in experiments (Fig.~\ref{fig:3}(A)), PNs showed a slower amplified response to transient inputs from ORNs, and LHN firing was more temporally refined, with the peak of their firing rate preceding that of presynaptic PNs, just as in experimental data (Fig.~\ref{fig:3}(B) Left). This rapid response of LHNs caused detection accuracy (${d^\prime}$) (see~\cite{Jeanne:2015di} and Appendix A2) for the ORN input to grow much faster to a larger maximum in LHNs than in PNs (Fig.~\ref{fig:3}(B) Right).

We also constructed and tuned homogeneous network models, in which PNs and LHNs are of the same type for comparison. In contrast, they showed suboptimal behaviors, such as the delayed firing of LHNs; therefore, ${d^\prime}$ of LHN rose more slowly and reached a lower maximum than that of PNs (Fig.~\ref{fig:3}(C)).

How do the different intrinsic properties of neurons contribute to the speed and high fidelity of LHN output? Since PNs and LHNs have opposite traits of differentiators and integrators, respectively, their effects can compensate for each other in the combined feedforward transformation of the ORN output. To analyze how the PN and LHN layers transform ORN inputs together, we computed how they amplify the power spectrum of ORN firing within a physiological frequency band ($\leq$ 100 Hz) with data from another longer simulation with the continuous current stimulus to ORNs. The results (Fig.~\ref{fig:3}(D)) showed that homogeneous networks with differentiator and integrator PNs/LHNs preferentially amplified higher or lower frequency components, respectively, whereas the heterogeneous network showed little distortion across the entire frequency range, demonstrating that PNs and LHNs compensated distortions introduced by each other.

We found that this mechanism also facilitated information transfer. We estimated (the lower bound of) mutual information (MI) between the input to ORNs and spike outputs of each layer~\citep{Borst:1999hw} (see also Appendix A2). In this way, we compared how much information in ORN firings pertaining to the input is transmitted to the output firing of PNs and LHNs. Specifically, we measured the information transfer from ORNs to PNs or to LHNs by computing a ratio of MIs, $I(\mbox{ORN input}; \mbox{PN or LHN output})/I(\mbox{ORN input}; \mbox{ORN output})$, respectively, where $I(X; Y)$ denotes MI between $X$ and $Y$. We found that information transfer to PNs closely matched that to LHNs in the heterogeneous network, whereas significant information loss was observed in homogeneous networks (Fig.~\ref{fig:3}(E)). In particular, the all-integrator PN/LHN case showed information loss specifically in the high-frequency band (Fig.~\ref{fig:3}(E), Right), indicating that large signal distortion in this regime (Fig.~\ref{fig:3}(D)) impaired information transfer. Interestingly, if we keep laminar heterogeneity but reversed the neuronal properties between PN and LHN, the network also shows stable power amplification and good information transfer (Appendix~\ref{appendix:rev}). These showed that laminar-specific intrinsic and functional properties of PNs and LHNs enabled nearly optimal information transfer between those neurons.

\subsection{Lamina-specific neuronal properties promote robust and stable signal propagation in deep FFNs}

\label{chap:deepffn}

We then investigated whether this mechanism can also enhance signal transmission in larger networks. For this purpose, we extended the AL network to a deep heterogeneous FFN model by adding more alternating layers of integrator or differentiator neurons. The deep FFN models had nine layers of 1,000 differentiator or integrator neurons in the AL network model, except for the input layer comprising differentiators. Again, each neuron was randomly connected to nine presynaptic neurons on average. Synaptic conductances and other parameters were the same as the AL network. The network parameters can be found in Appendix Table A1, A2.

We simulated how a packet of spikes, injected into the input layer, propagates through subsequent layers~\citep{Diesmann:1999ina, vanRossum:2002vz, Reyes:2003gb, Kumar:2008ir, Kumar:2010dv, Kremkow:2010gx, Moldakarimov:2015gb, Joglekar:2018gq}. Spike generators randomly sampled in total $\alpha$ spike times from a normal distribution with variance $\sigma^2$ and forced the input layer neurons to fire at the spike times, in addition to noisy spontaneous firing.

\begin{figure}[t]
    \centering
    \includegraphics[width=1.0\textwidth]{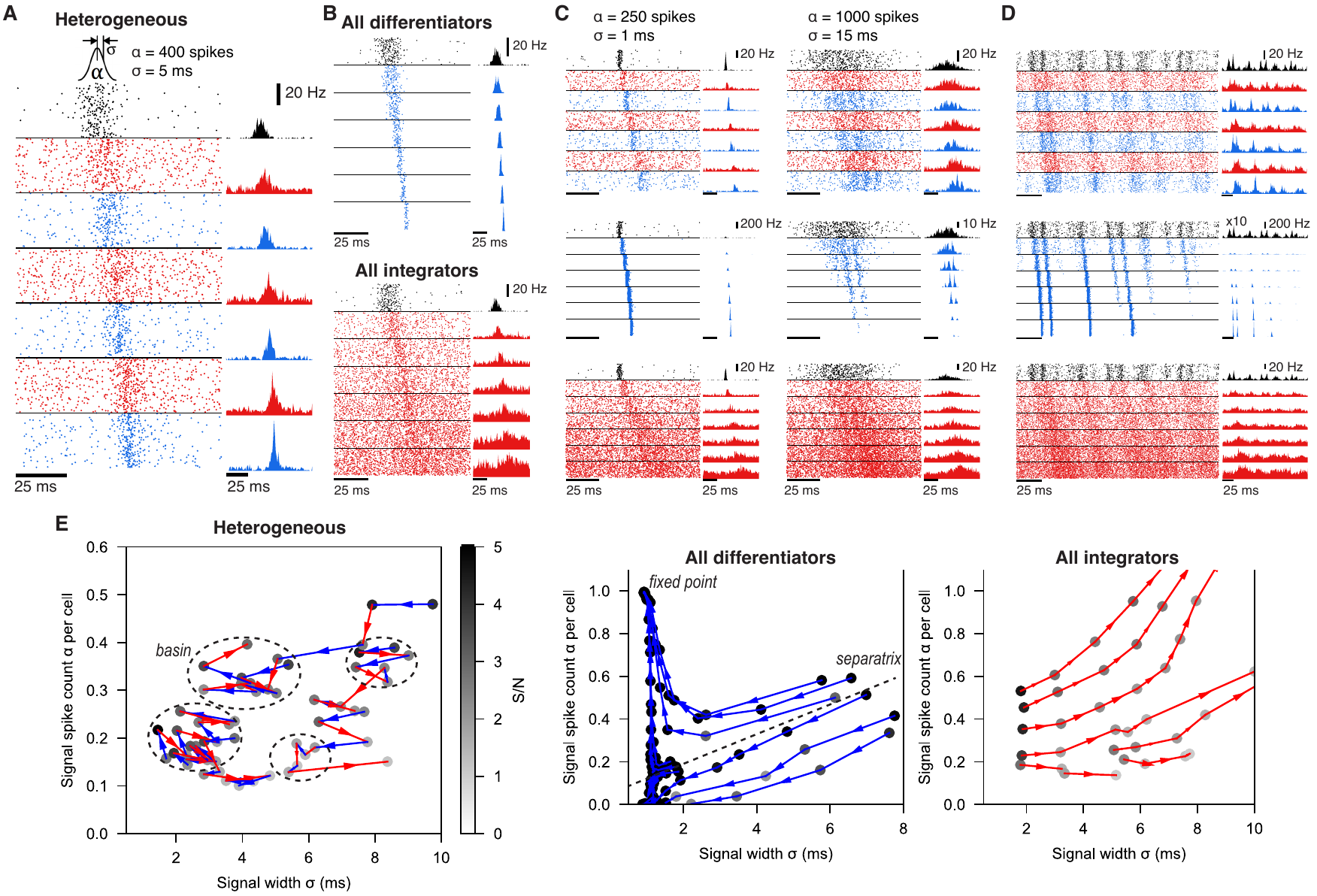}
    \caption{\textbf{Lamina-specific neuronal properties robustly stabilize spike signal propagation in deep FFNs.} (A) Propagation in a heterogeneous network. Inset on top is the Gaussian distribution of spikes, evoked in the input layers (black). Dots (Left) and histograms (Right) are spikes and their firing rates, respectively. In all figures, blue and red represent differentiators and integrators, respectively (B) Firing in homogeneous networks with only differentiators (Top) and integrators (Bottom). (C) Propagation of signals with different spike count ($\alpha$) and width ($\sigma$). (D) Network firing with continuous noise current in the input layer. In the middle row (blue; all differentiator), the input layer firing rate is multiplied by 10 for clarity. (E) Analysis of signal transformations underlying stable propagation in the ($\sigma$, $\alpha$) space. Each trajectory is formed by connecting ($\sigma$, $\alpha$) of a propagating signal (dots) between adjacent layers, starting from the second layer output. Shade of each dot is the signal-to-noise ratio (S/N) and only points with S/N>1 are shown. Dotted circles mark “basins” (Fig.~\ref{fig:1}(B)) where any propagating signal stays for $\geq$5 layers. A dotted line in the Middle panel is an approximated separatrix between trajectories toward a fixed point and dissipation (Fig.~\ref{fig:1}(A)). All models have 9 layers and the first 7 layers are shown in a-d for clarity.}
    \label{fig:4}
\end{figure}

We found that the spike signals stably propagated in this network, whereas homogeneous networks, with only differentiators or integrators, showed opposing results (Fig.~\ref{fig:4}(A,B)): In the all-differentiator network, the evoked spike signal became increasingly synchronized and propagated as layer-wide synchronized spikes. In contrast, in the all-integrator network, the evoked spike signal became broader and less synchronized, until it eventually dissipated into spontaneously firing spikes (Fig.~\ref{fig:4}(A) Right). Stable propagation in the heterogeneous network was decidedly robust over a wide range of input signals with diverse temporal width ($\sigma$) and the total number of spikes ($\alpha$) (Fig.~\ref{fig:4}(C) Top). Conversely, the all-differentiator network exhibited a clear preference for sharply synchronized spikes, while signals gradually dissipated into spontaneous activity in the all-integrator network (Fig.~\ref{fig:4}(C) Middle-Bottom). Therefore, when tested with input signals with diverse ($\sigma$, $\alpha$), the heterogeneous network showed the best performance in signal propagation (Appendix Fig. A2), and this result did not significantly change with additional feedforward inhibition in the deep FFN (Appendix Fig.~A2).

Furthermore, we examined the case in which the input layer spiked with the dynamically fluctuating firing rate, due to dynamical, stochastic current injection (the same OU process as the AL network, Equation~\ref{eq:ou2}, except that $\mu_{input}$=25 $\mu$A/c$\mbox{m}^2$ and $\sigma_{input}$=12.5 $\mu$A/c$\mbox{m}^2$. See Appendix Table A3 for the other parameters). This continuous signal propagated in the heterogeneous network with many conserved features, whereas significant signal distortion and loss were again observed in the homogeneous networks (Fig.~\ref{fig:4}(D)). Note that propagation of dynamical input features indicates superior information transfer in a heterogeneous network, compared to homogeneous ones.

Again, the robust and stable signal propagation was possible by the distortion-compensating input/output transformations by neighboring layers with distinct neuron types. To demonstrate this, we analyzed trajectories of propagating signals in the ($\sigma$, $\alpha$) plane~\citep{Diesmann:1999ina, Kumar:2008kt} (Fig.~\ref{fig:4}(E)), a simple version of the signal space that we previously discussed (Fig.~\ref{fig:1}). In the heterogeneous network, each layer transformed an incoming signal into a different, sometimes nearly opposite or complementary direction in the ($\sigma$, $\alpha$) plane than those transformed by its pre- and postsynaptic layer, which prevents the formation of a uniform flow. Therefore, a propagating signal cannot run away and is confined to a small region (basin), corresponding to stable propagation (Fig.~\ref{fig:4}(E) Left). However, in homogeneous networks, all layers perform similar transformations and drive propagating signals rapidly toward a fixed point of sharp synchronization or dissipation (Fig.~\ref{fig:4}(E) Middle, Right). Notably, in most of the ($\sigma$, $\alpha$) plane, transformations in those two networks are in nearly opposite directions: In the all-differentiator network, $\sigma$ and $\alpha$ both tend to decrease (Fig. 4e Middle), because sharply correlated spikes are the preferred input of the neurons, while $\sigma$ and $\alpha$ increase in the all-integrator network (Fig.~\ref{fig:4}(E) Right). In the heterogeneous network, those two different transformations are performed by neighboring layers to minimize overall signal distortion and boost information transfer. In summary, the distortion compensation mechanism by distinct neuron types can protect a propagating signal from undergoing a loss or distortion regime in the signal space, while supporting the robust and stable transmission.

\subsection{Signal amplitude-dependent propagation latency in the heterogeneous FFN}

\begin{figure}
    \centering
    \includegraphics[width=1.0\textwidth]{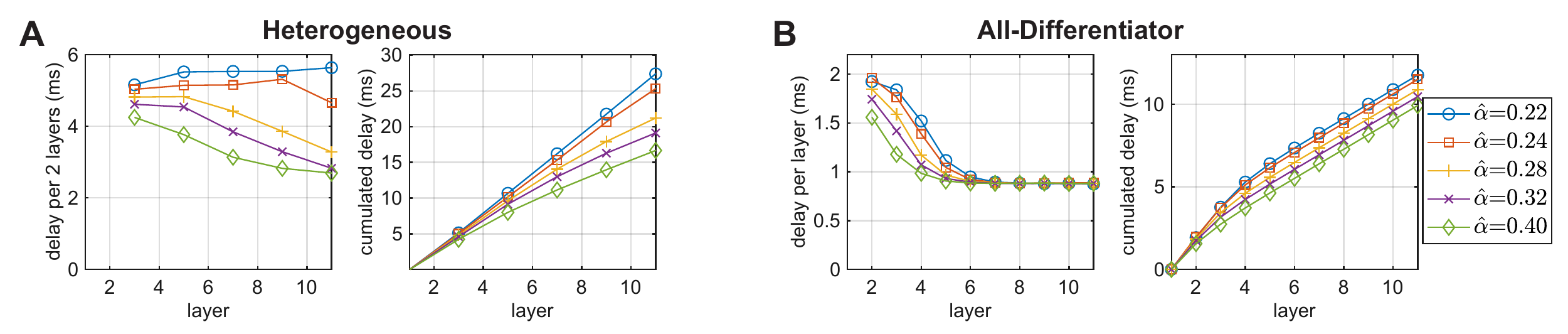}
    \caption{Propagation latency of spike signals in the (A) heterogeneous and (B) all-differentiator FFN. We used the input signals with $\sigma=1$ ms for different firing rate, which therefore had the different numbers of spikes per cell in the input layer, $\hat{\alpha}$. The delays were computed as the differences of the median spike time $t_c$ of the signals between layers (see Appendix~\ref{appendix:data_ana}).}
    \label{fig:delay}
\end{figure}

While many studies have investigated the stable propagation of spike signals, the question of how rapidly the signals propagate through layers has received relatively less attention. However, millisecond-level signal transmission latencies in neural circuits are known to play crucial roles~\citep{VanRullen:2005lq, Chechik:2006qr}, e.g. in sound localization by birds and mammals using interaural time differences~\citep{grothe2010mechanisms}. The propagation latency depends on the properties of synaptic connections, such as dendrite length, myelination, etc. In deep FFNs, the features of a propagating signal can also determine its propagation latency, since a transmission latency from one layer to another depends on how rapidly presynaptic spikes trigger spiking in the postsynaptic neurons, a property governed by their intrinsic mechanisms.

To examine this, we simulated our deep FFNs, now with more neurons (10,000) in each layer (and reduced the connection density correspondingly to keep each neuron randomly connected to 9 presynaptic neurons on average) for better statistical testing (Fig.~\ref{fig:delay}). The results show that in the heterogeneous FFN, the propagation latency in deep layers varied with the input signal amplitude $\alpha$ (Fig.~\ref{fig:delay}(A)). In the case of all-differentiator FFN, the propagation delay varied only for a few initial layers and quickly converged in deeper layers (Fig.~\ref{fig:delay}(B)). In the case of the all-integrator FFN, the quick dissipation of spike signals (Fig.~3(B-D)) made the measurement of signal latency meaningless.

This result suggests that lamina-specific cell types in deep FFNs can naturally lead to neurons using both of the spike rate and time coding for an input signal \citep{Panzeri:2010fk, Ratte:2013fg}, especially in deeper layers. Therefore, deeper layers can develop more complex and expressive representations of network inputs. It will be of interest to further investigate the emergent coding properties in the deep spiking neural network, as a model of the neural systems, and their cellular basis.

\section{Discussion}
Diversity of cell types is one of the distinctive characteristics in neural systems, and its functional characterization is the subject of ongoing experimental investigations (e.g., \cite{gouwens2019classification}). Integrating information about cell types and their intrinsic properties with network connectivity should be a pivotal research question to develop a holistic understanding of how spike signals propagate in neural circuits. However, the diversity of cellular properties is one of the most neglected elements in theoretical neural network studies.

We focused on functionally distinct cell types due to different voltage-dependencies of $\mbox{K}^+$ channels, which can arise from diverse expression patterns of low-threshold $\mbox{K}^+$ channels~\citep{Higgs:2011eu,hu2007m,Svirskis:2002br}. However, other neuronal mechanisms that affect the integrative cellular property can play similar roles, such as morphology~\citep{mainen1996influence}, inactivation of $\mbox{Na}^+$ channels~\citep{Arsiero:2007bk, Mensi:2016he}, h-channels~\citep{Kalmbach:2018bp}, the high-conductance state~\citep{destexhe2003high, Prescott:2008bg}, etc. Furthermore, synaptic and circuit mechanisms known to operate as integrators or differentiators can be organized by a similar principle. For example, short-term synaptic depression and facilitation can act as high- or low-pass filters, respectively~\citep{Izhikevich:2003tp}, and inhibition can limit the integration time window for incoming inputs and promote temporal fidelity of neuronal responses~\citep{Pouille:2001hw}. Our hypothesis predicts that integrator neurons, such as PNs, tend to have synapses with short-term depression~\citep{Kazama:2009ho} whereas differentiators, such as LHNs, have facilitating synapses.

Jeanne and Wilson compared spike signal transfer from thalamocortical to cortical layer IV neurons to that between PNs and LHNs~\citep{Jeanne:2015di}. Likewise, we further propose that these theoretical mechanisms can be applied to the thalamocortical loop and cortico-cortical feedforward projections, where spike signals propagate through multiple types of principal neurons that are different in size, morphology, ion channel expressions, etc. for each layer. Stable signal propagation in an FFN has been extensively studied in this context~\citep{Diesmann:1999ina, Joglekar:2018gq, Kremkow:2010gx, Kumar:2008ir, Kumar:2010dv, Moldakarimov:2015gb, Reyes:2003gb, vanRossum:2002vz, Vogels:2005fc}. However, proposed models so far were often successful only with a limited range of input signals given fixed model parameters, although some precisely tuned models can handle a diverse range of inputs~\citep{Kumar:2008ir, Kumar:2010dv, vanRossum:2002vz, Vogels:2005fc}. In this study, we proposed a novel approach to this problem, based on an information-theoretic perspective, pointing out that an assumption of a single cell type in a network can result in accumulated signal distortion.  Introducing multiple cell types with lamina-specific neuronal properties can circumvent this problem, exemplified by stable propagation of a dynamical spike signal in our model. Given the prevalence of diverse cell types in many neural systems, our work presents a clear case that lamina-specific cell types are critical to understanding network functions.

If pre- and postsynaptic neurons have differences in the intrinsic integrative properties, it inevitably causes a mismatch in their input/output transformations, and we suggest that this phenomenon can be a signature of optimal information transfer. A previous study suggested a different interpretation that the postsynaptic layer filters out a substantial fraction of information encoded by an input layer if such a mismatch is present ~\citep{blumhagen2011neuronal}. PNs and LHNs in the \textit{Drosophila} AL network also have mismatching input/output transformations due to differences in their intrinsic properties. However, our model showed that the information transfer from PNs to LHNs was nearly optimal, instead of LHNs filtering down a significant fraction of the information carried in the PN firing. Likewise, distinct stimulus encoding schemes between pre-/postsynaptic neurons, ubiquitously found in many neural systems, can be a mechanism for boosting information transfer across the entire network. Although not identical to our case, there is an analog in artificial deep neural networks, the \textit{Residual Network}~\citep{he2016deep}, which significantly outperforms conventional deep networks consisting of homogeneous convolutional layers. The residual network uses two types of transmission functions (a parameterized mapping $f(x)$ and a shortcut connection $x+f(x)$), which allows gradients to backpropagate across deep layers without vanishing or exploding~\citep{he2016deep}.

Performing compensatory signal transformation is a widely used strategy in information-theoretic algorithms for optimizing information transfer with limited bandwidth, such as water-filling~\citep{Gallager:1968tl}. In this study, we have demonstrated how this scheme operates in FFNs when lamina-specific neuron types have different intrinsic properties. Notably, functionally different cell types within a layer can also be explained by the maximization of information transmission~\citep{kastner2015critical}. Therefore, we suggest that the commonly observed diversity of cell types in neural circuits is essential to achieve optimal information transmission.


\section*{Broader Impact}


Many efforts have been paid to understand the critical components of highly cognitive systems like the human brain. Studies have argued for simulations of large brain-scale neural networks as an indispensable tool~\citep{de2010world}. Still, they almost always fail to consider cellular diversity in the brain, whereas more and more experimental data are revealing its importance. Our computational study suggests that  heterogeneity in neuronal properties is critical in information transfer within a neural circuit and it should not be ignored, especially when the neural pathway has many feedforward layers.

For deep-learning research, our work also provides a new insight for neural architecture search (NAS)~\citep{elsken2019neural}. The search space of existing NAS methods are mainly (1) the combination of heterogeneous layers to form an entire network; (2) the combination of heterogeneous activation functions to form a cell. However, our work suggests a novel, computationally efficient strategy, that is searching for a block structure consisted of several layers (In our case, the block is composed of a integrator layer followed by a differentiator layer). On the one hand, the block should boost stable propagation of input signals into deep layers. Hence, divergence of inputs will remain detectable in the output layer at the initial phase of learning, which is suggested to accelerates the training of very deep networks~\citep{schoengloz2016deep, srivastava2015training}. On the other hand, there is also extra freedom of searching the block structure that does not suffer from vanishing/exploding backpropagation gradients (like a residual block~\citep{he2016deep})


Our deep FFN models are proof-of-concept and lack many other neural circuit mechanisms that can affect signal propagation in spiking neural networks, as we discuss in Section~\ref{chap:related_work}, although we did find that an additional component, feedforward inhibition, did not significantly change the results (Appendix Fig.~\ref{fig:s1},\ref{fig:s2}). Our study suggests that the cooperation between different types of neurons is vital for promoting signal processing in large-scale networks. It also suggests investigating the roles of heterogeneous neuronal properties in other problems such as sensory coding, short-term memory, and others, in the future studies.


\section*{Acknowledgements}

The authors declare no conflict of interest. We thank James Jeanne for helpful discussions and for sharing experimental data. We also thank Steven Prescott, Mario Negrello, Jihwan Myung, and Steven Aird for reading an earlier version and providing useful feedback. This work was supported by funding from the Okinawa Institute of Science and Technology Graduate University. SH was also supported by Japan Society for the Promotion of Science, KAKENHI Grant Number 15K06725.

\bibliography{paper}
\bibliographystyle{apalike}

\newpage
\begin{appendices}
\renewcommand{\thesection}{A\arabic{section}}
\renewcommand{\thetable}{A\arabic{table}}
\renewcommand{\thefigure}{A\arabic{figure}}

\setcounter{figure}{0}
\setcounter{equation}{0}
\setcounter{table}{0}

\section{Dynamical spiking threshold of the differentiator neurons}

\label{ALDIFINT}
\begin{figure}
    \centering
    \includegraphics[width=0.6\textwidth]{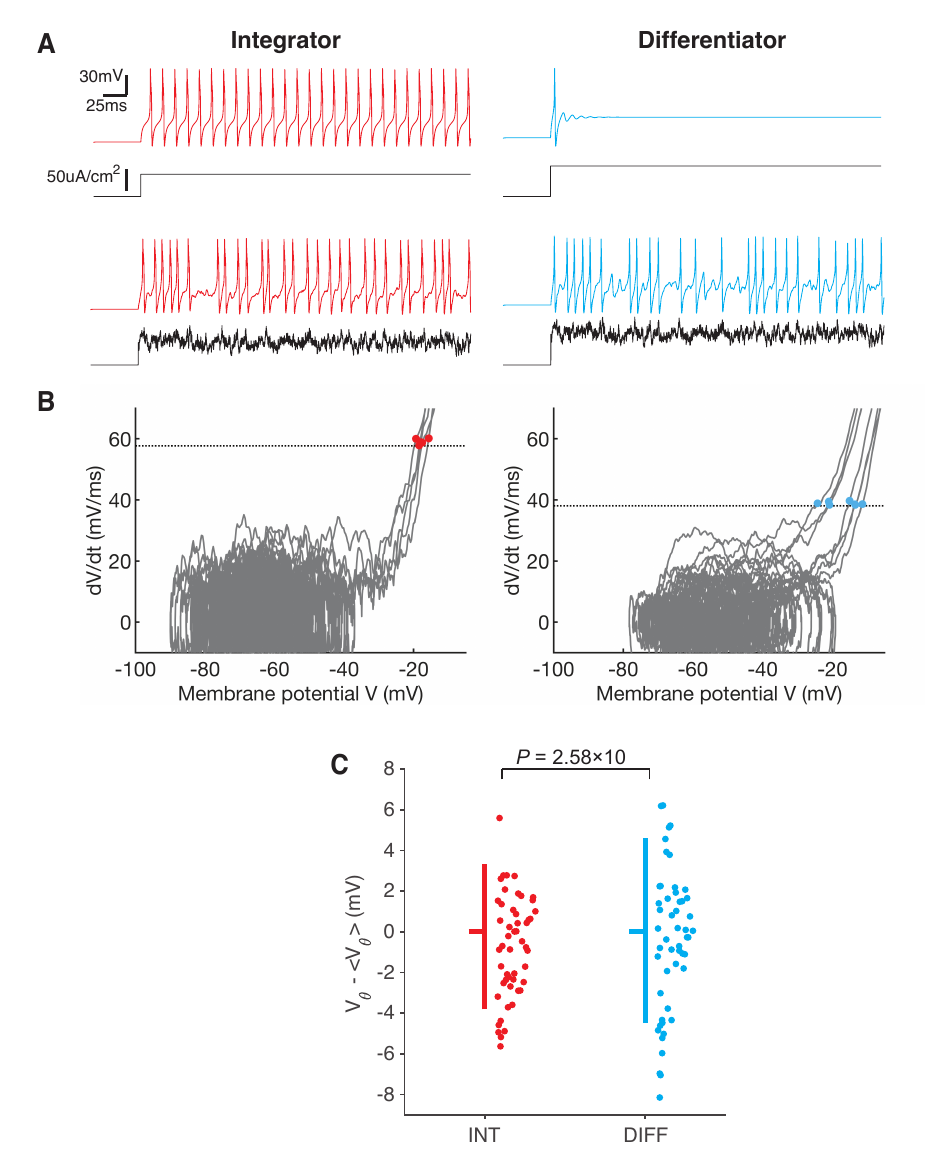}
    \caption{\textbf{Intrinsic properties of conductance-based model neurons control dynamicity of spiking thresholds.} (A) Membrane potential response (color) to constant or fluctuating current injection (black). (B) Example membrane potential $V$ vs. $dV/dt$ in two neurons, based on simulation data in Fig.~\ref{fig:3}(A,B). Data from one trial are shown (gray). Dotted lines represent $[dV/dt]_{\mbox{min}}$, the minimal $dV/dt$ for spiking, and colored dots are threshold-crossing points. (C) Spread of membrane potentials at crossing points, $V_\theta$, from the average. Vertical bars span from 10\% to 90\% quantiles, and notches are at medians. Data are the same as B, and only 50 samples (dots) are shown for clarity.}
    \label{fig:2}
\end{figure}

The dynamic threshold is a neural property where firing not only depends on the membrane potential but also on its temporal change, which is crucial for neural sensitivity to input fluctuations~\citep{Prescott:2008bg,Azouz:2003vb,Hong:2008gg}. An experimental study showed that LHNs have dynamic thresholds~\citep{Jeanne:2015di}, and differentiator neurons are also known to have the property. Therefore, we used differentiator neurons for modeling LHNs, and integrators with for PNs in the AL network. To demonstrate their difference in the spiking threshold, we estimated a minimal rate of membrane potential change, $ [dV/dt]_{\mbox{min}} $, that preceded spikes, but not subthreshold fluctuations from our simulation data, which corresponds to the minimal inward current required for spiking (Fig.~\ref{fig:2}(B)). Then, the threshold voltage, $V_\theta$, at $dV/dt \approx [dV/dt]_{\mbox{min}}$, was significantly more distributed in differentiators (Integrator: STD[$V_\theta$]=2.76$\pm$0.08 mV, Differentiator: 3.33$\pm$0.11 mV; \textit{P}=2.58$\times10^{-5}$, \textit{F}-test; Fig.~\ref{fig:2}(C)), suggesting that differentiators can generate enough inward current to generate a spike across a broader range of membrane voltages than integrators, an indication of a more dynamical spiking threshold. Therefore, we used differentiator neurons, with the low-threshold $\mbox{K}^+$ channel, for modeling LHNs, and integrators with the high-threshold channel for PNs in the AL network.

\section{Data analysis}
\label{appendix:data_ana}
In the AL network model case, ${d^\prime}$, a measure for signal detection, was computed in the same way as in~\citep{Jeanne:2015di}:
\begin{align}
	{d^\prime}=\frac{\mu_{stim}-\mu_0}{\sqrt{(\sigma_{stim}^2+\sigma_0^2)/2}}
\end{align}
where $(\mu_{stim}, \sigma_{stim})$ and $(\mu_0, \sigma_0)$ are the (mean, STD) of spike count at a given layer, computed with 80 ms-long overlapping temporal windows in the stimulated and non-stimulated condition, respectively. For each layer, we computed ${d^\prime}$ of all the cells and plotted their median in Fig.~2(B,C).

Power spectra for Fig.~2(D) were evaluated by applying the MATLAB function \texttt{pmtm} with a 20-ms time window on spike trains formed with 1-ms time bins. Mutual information in Fig. 3E were computed by a Gaussian channel approximation~\citep{Borst:1999hw}: We first reduced the dimensionality of a population spike trains at each layer, by using principal component analysis (PCA). Since the first PCA component was always dominating, we projected the population spike trains to this component to form a one-dimensional “population response” time series. With the Fourier transformation of the stimulus and population response, $S(\omega)$ and $R(\omega)$, we estimated a kernel $K(\omega) = <R^*(\omega)S(\omega)>/<R^*(\omega)R(\omega)>$, and computed a reconstructed stimulus and noise via $Sr(\omega) = R(\omega)K(\omega)$ and $N(\omega) = S(\omega)-Sr(\omega)$. The mutual information per each frequency bin was then computed by
\begin{align}
    	I(S(\omega);R(\omega))=\log_2 (1+SNR(\omega)), \quad SNR(\omega)=\|S_r (\omega)\|^2 / \|N(\omega)\|^2.
\end{align}
With this, we computed the information transfer (Fig. 3E) by $$ T_X(\omega) = I(S(\omega)_\text{ORN};  R(\omega)_X)/I(S(\omega)_\text{ORN}; R(\omega)_\text{ORN}),$$ where $X$ is PN or LHN.

In the deep FFN, we computed $(\sigma, \alpha)$ for spikes from each layer using a custom algorithm that estimates $(\sigma, \alpha)$ in the presence of additional spontaneous firing. We first computed the baseline spontaneous firing rate $\nu_0$ at each layer by averaging the firing rate obtained from the same model with no input. The firing rate curve was computed by histogramming spike times in this layer with a 0.1-ms time bin and by smoothing it with a 3-step moving average. Then, we evaluated a least-square fit of $\nu(t)$ to $\nu_{fit}(t) = \nu_0 + \nu_1 \exp(-(t-t_c)^2/2\sigma^2)$. $\alpha$ was estimated by counting the spikes in the [$t_c - 3\sigma$, $t_c + 3\sigma$] window. From the goodness of fit, $R^2 = 1 - < (\nu(t)- \nu_{fit}(t))^2> / \mbox{Var}[\nu(t)]$, we evaluated the signal-to-noise ratio, $S/N = R/(1-R^2)^{1/2}$ (Fig.~\ref{fig:4}(E)).

All analysis was performed by custom codes written in MATLAB 2016b (MathWorks, MA) and Python. All the models are publicly available at \url{https://github.com/FrostHan/HetFFN-}. The  datasets generated during and/or analyzed during the current study and analysis codes are available from the corresponding author upon reasonable request.

\section{Reversed heterogeneous FFNs and intermediate $\beta_w$}
\label{appendix:rev}

In the AL network, we modeled PNs as integrators and LHNs as differentiators according to the experimental findings~\citep{Jeanne:2015di}. As alternative cases, we simulated the AL network of homogeneous PNs and LHNs with intermediate $\beta_w$=-10 mV, and the same network with the properties of PNs and LHNs reversed (i.e. PNs and LHNs are differentiators and integrators, respectively). The results of homogeneous AL network with intermediate $\beta_w$ are shown in Fig.~\ref{fig:rev}({(A-C)}) Left and D. Compared with the original heterogeneous AL model (Fig.~\ref{fig:3}), the homogeneous AL network performs worse in terms of accuracy and information transfer (Fig.~\ref{fig:rev}({(B-C)} Left), and shows less stable power amplifcation (Fig.~\ref{fig:rev}({(D)}). The reversed AL network shows the stable power amplification and good information transfer (Fig.~\ref{fig:rev}({(C)}) Right, and (D)). However,  ${d^\prime}$, accuracy of the ORN input detection, is suboptimal (Fig.~\ref{fig:rev} (B) Right; dots are lower than solid lines), since ORNs fire sparsely with strong differentiator characteristics \citep{nagel2011biophysical} and, therefore, integrators can be better suited for their postsynaptic cells.

We further extended these two kind of network models to deep FFNs (Fig.~\ref{fig:intmedraster}). In the homogeneous FFNs with intermediate $\beta_w$ = -7 mV, -10 mV and -12 mV, a relatively weak and asynchronous spike signal dissipated in deep layers (Fig.~\ref{fig:intmedraster} A-C), while a relatively strong and synchronous one tended to diverge (Fig.~\ref{fig:intmedraster} E-G). In contrast, the reversed heterogeneous network showed robust and stable signal transmission (Fig.~\ref{fig:intmedraster} D, H), similar to our original heterogeneous FFN model. Therefore, stable signal propagation can be achieved only by proper demodulation of signal distortions between adjacent layers, regardless of the order of integrator / differentiator. In contrast, if an FFN contains only homogeneous layers, the signal propagates with accumulated distortion (amplification/dicrease in $\alpha$/$\sigma$) into deep layers.

\section{Sensitivity to $\beta_w$ in deep heterogeneous FFNs}
\label{appendix:sensitivity}

$\beta_w$ for integrator ($5$ mV) and differentiator ($-19$ mV) neurons in the deep heterogeneous FFNs, together with other hyper-parameters, were from the AL network model. Will changing $\beta_w$ impair stability of signal propagation in the heterogeneous network? We show a sweep of $\beta_w$ for integrator ($3, 5, 7$ mV) and differentiator ($-20, -19, -18$ mV) neurons in the deep heterogeneous FFN (Fig.~\ref{fig:sweep}).

Notably, varying $\beta_w$ for integrator neurons did not significantly affect the propagation property, while a different value of $\beta_w$ for differentiators changed the property. This is because the mean spike threshold increased significantly when $\beta_w$ of differentiator became smaller~\citep{Ratte:2013fg}. Therefore, the range of input ($\alpha, \sigma$) that can stably propagate reduced when $\beta_w=-20$ mV for differentiators (Fig.~\ref{fig:sweep} A, D, G), and enlarged when $\beta_w=-18$ mV for differentiators (Fig.~\ref{fig:sweep} C, F, I). However, we could compensate for the change in the spike threshold of differentiators, which was introduced by varying $\beta_w$, by increasing or decreasing the synaptic conductance from integrators to differentiators. Fig.~\ref{fig:sweep} (J-K) shows that a similar propagation property to the original heterogeneous FFN model can be achieved by doing so.

As we discussed in Section~\ref{chap:deepffn}, the robust and stable signal propagation is due to the distortion-compensating input/output transformations by neighboring layers with distinct neuron types. As long as this mechanism is not profoundly compromised, spike signals should be able to transmit stably and robustly.

\begin{figure}
    \centering
    \includegraphics{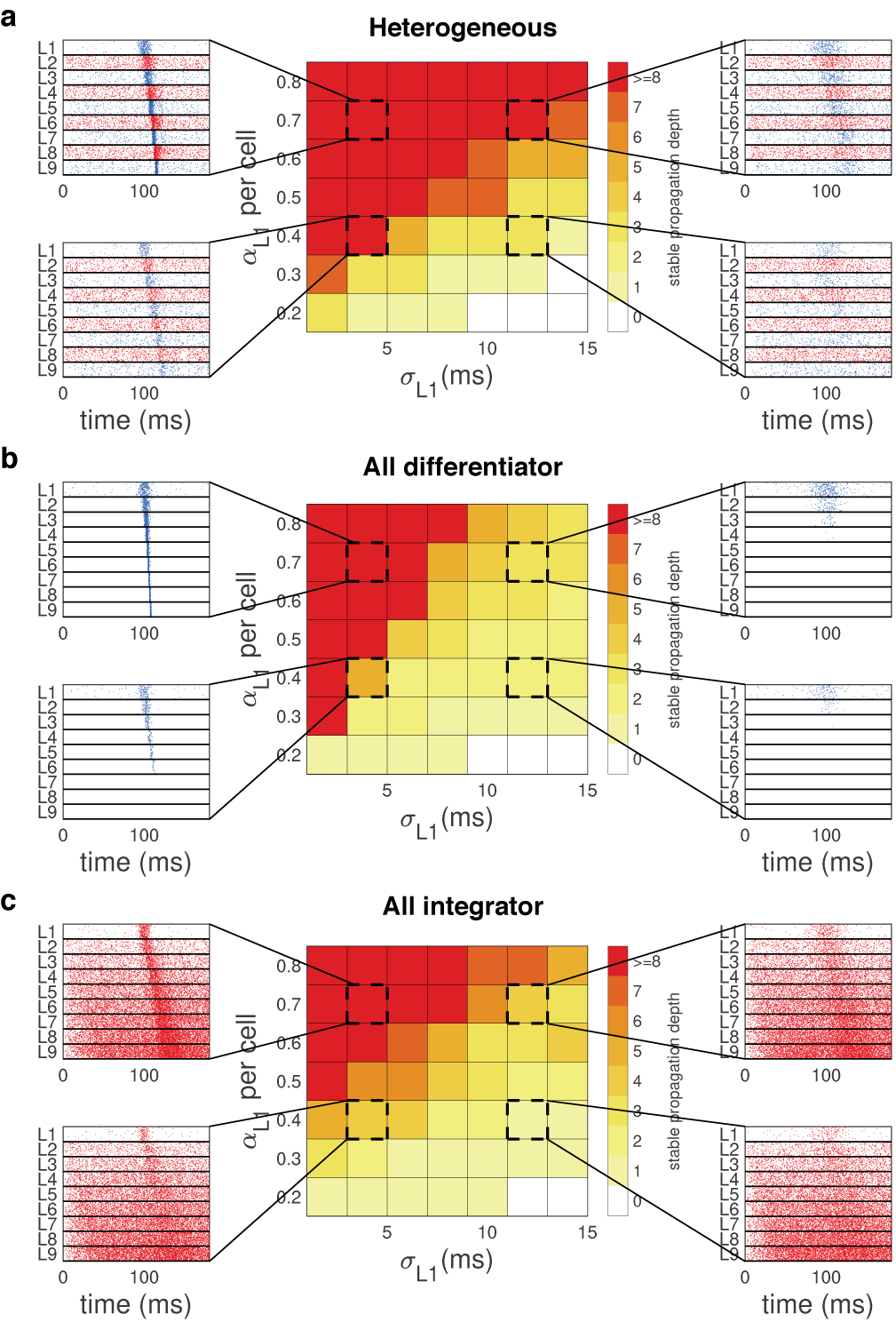}
    \caption{\textbf{Propagation of spike signals with diverse width ($\sigma$) and number of spikes ($\alpha$) in the heterogeneous (A), all-differentiator (B), and all-integrator network (C).} Each network has nine layers of 5,000 neurons (see Table A3 for parameters). Color in the middle column represents propagation depth, computed by numbers of layers (except the input layer) into which spike signals propagate. Propagation is considered stopped if the estimated $\alpha$ is lower than 0.05n or larger than 3n for a layer and its corresponding postsynaptic layer, where n=5,000 is the group size. Side insets are example raster plots for parameters marked by dotted squares in the middle, showing spikes from 10\% of neurons at each layer for clarity.}
    \label{fig:s1}
\end{figure}

\begin{figure}
    \centering
    \includegraphics{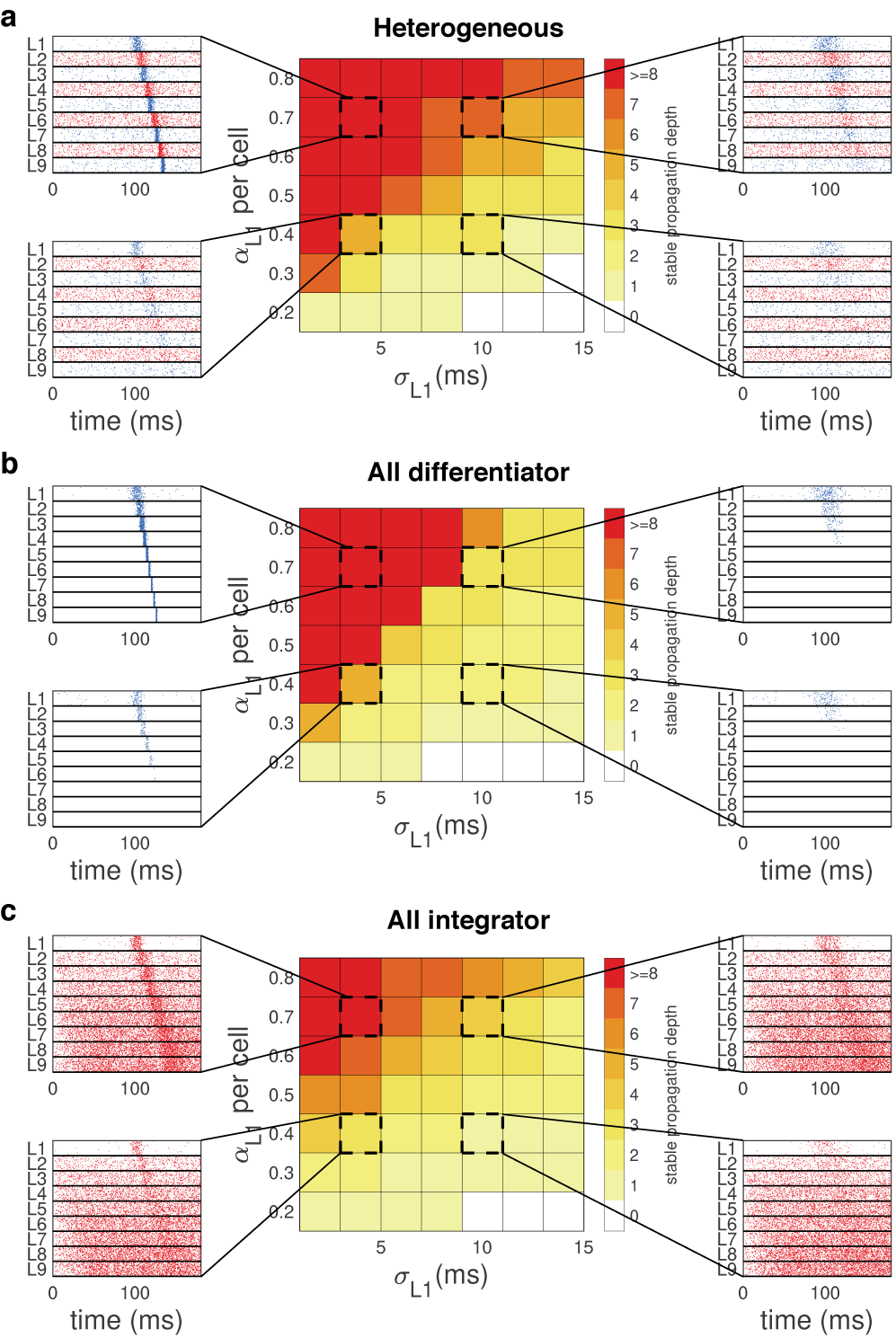}
    \caption{\textbf{The same figures as Fig.~\ref{fig:s1}, but using FFN models with feedforward inhibition.} Again, each network has nine layers of 4,000 PN-like or LHN-like excitatory neurons, combined with 1,000 inhibitory neurons that receive excitatory inputs from a previous layer and inhibit excitatory neurons in the same layer. Inhibitory cells were also based on the Morris-Lecar model with $\beta_w$ = -15 mV while different $\beta_w$ did not cause any significant change in our conclusion. The reversal potential of inhibitory synapses was $E_{syn}$ = -90 mV and the conductance was 200 $\mu$S/c$\mbox{m}^2$. Also, we added a synaptic delay of 2 ms for all connections. Other parameters were the same as those for Fig.~\ref{fig:s1}. In all panels, we plotted spikes from 10\% of excitatory neurons at each layer for clarity.}
    \label{fig:s2}
\end{figure}

\begin{figure}
    \centering
    \includegraphics[width=0.6\textwidth]{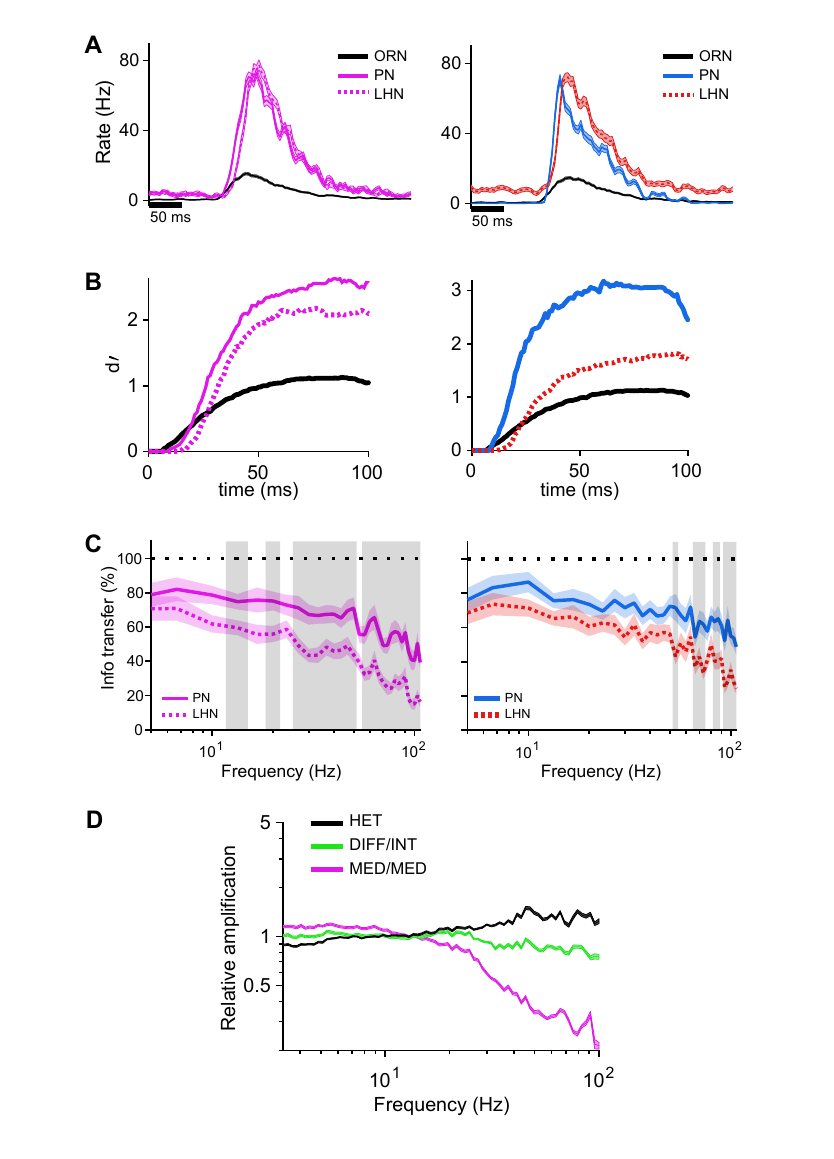}
    \caption{(A-C) Firing rates (A), ${d^\prime}$ (B), and information transfer (C) for the homogeneous AL network with $\beta_w$=-10~mV (left) and reversed heterogeneous model (right). (D) Power amplification of the original (black), reversed (green), and $\beta_w$=-10~mV network model. Blue: Differentiator, Red: Integrator, Magenta: $\beta_w$=-10~mV. Shade in C: \textit{P}<0.01.}
    \label{fig:rev}
\end{figure}

\begin{figure}
    \centering
    \includegraphics[width=1.0\textwidth]{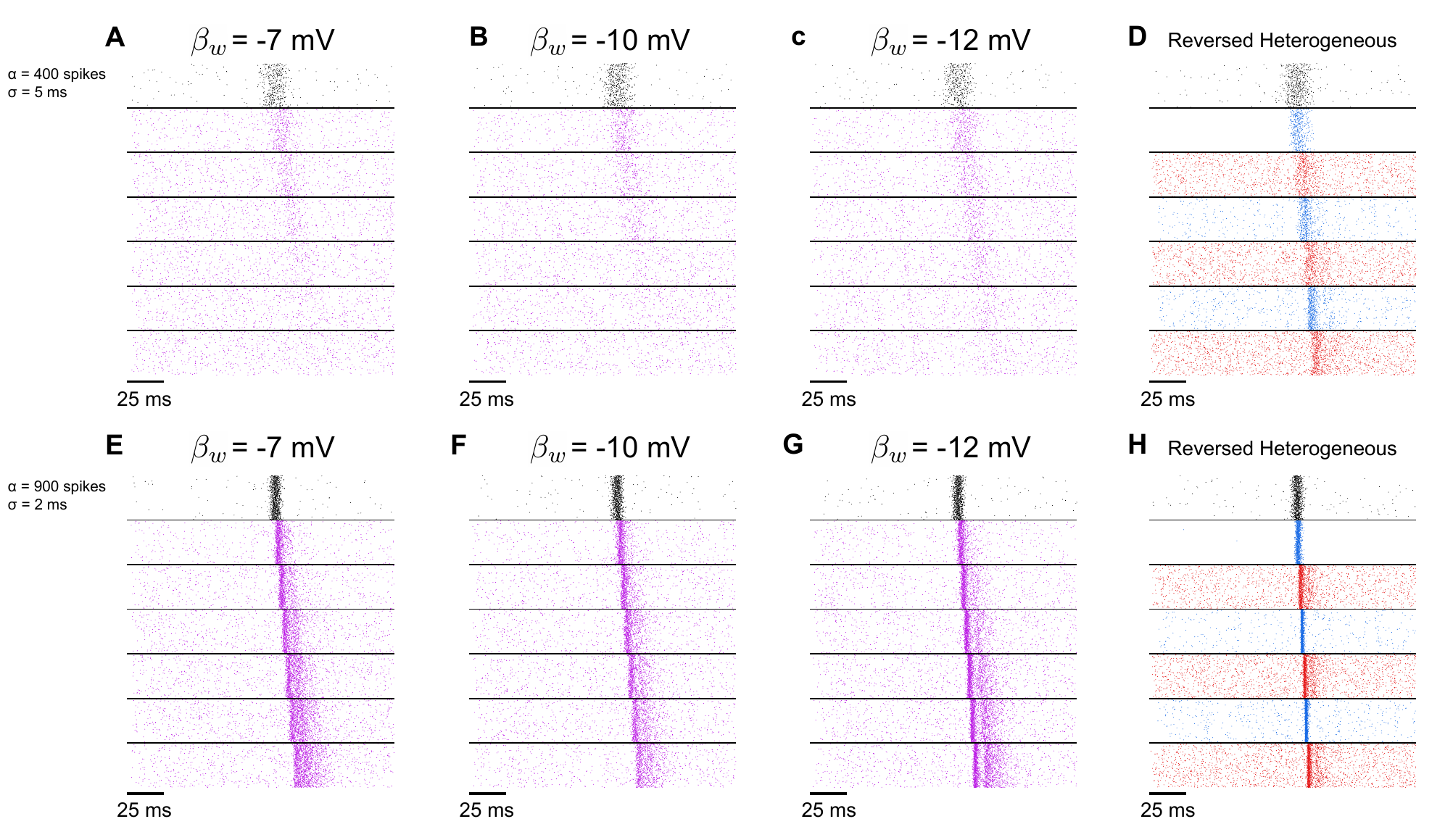}
    \caption{Raster plot of deep homogeneous FFNs with intermediate $\beta_w$, and the deep reversed-heterogeneous FFN. The input spike signal is featured with ($\alpha$=400~spikes, $\sigma$=5~ms) for the first row (relatively weak and asynchronous input signal) and ($\alpha$=900~spikes, $\sigma$=2~ms) for the second row (relatively strong and synchronous input signal).}
    \label{fig:intmedraster}
\end{figure}


\begin{figure}
    \centering
    \includegraphics[width=1.0\textwidth]{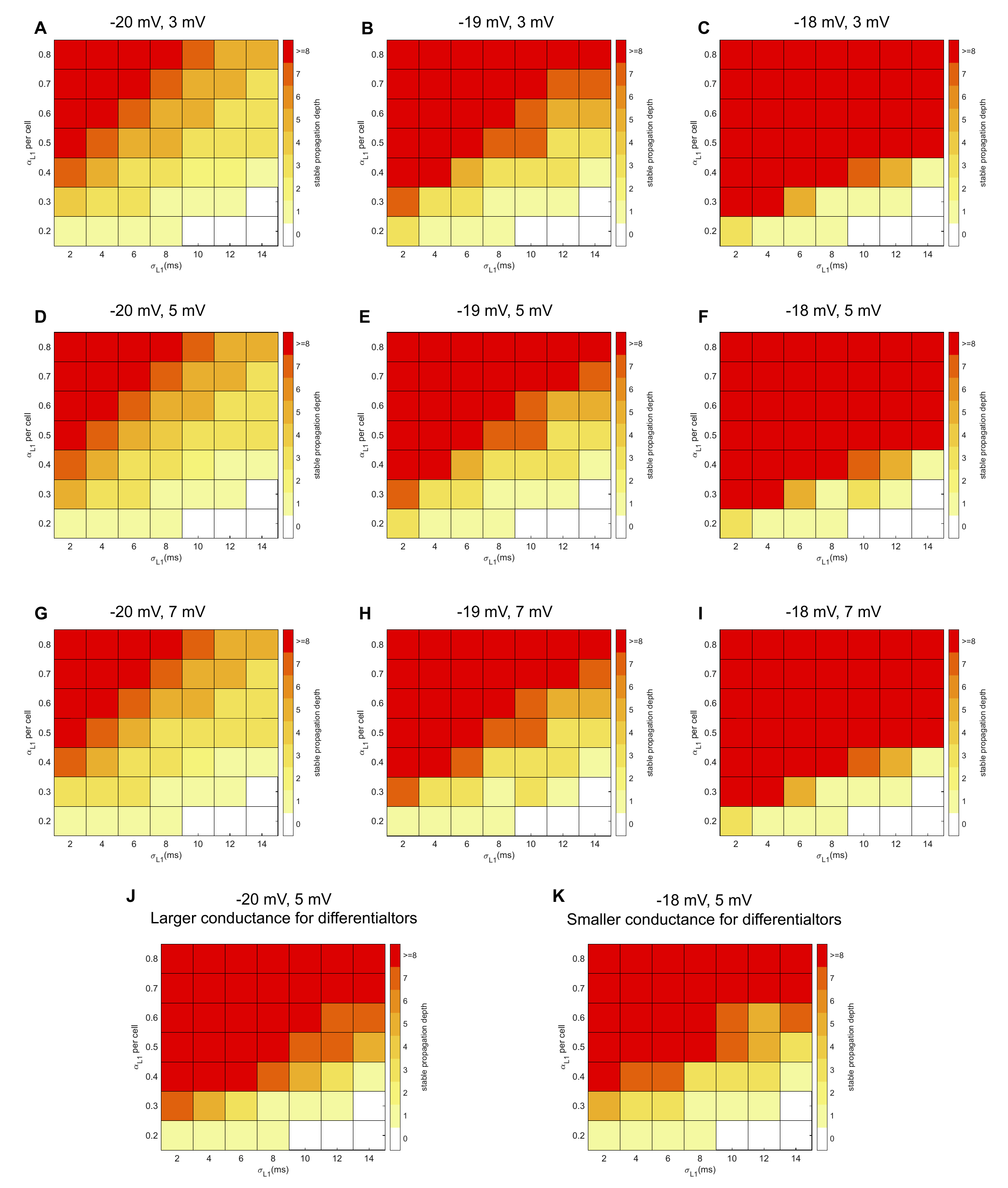}
    \caption{(A-I) Propagation of spike signals with diverse width ($\sigma$) and number of spikes ($\alpha$) in the heterogeneous networks with changed values of $\beta_w$ for differentiator (the first voltage) and integrator neurons (the second voltage) while keeping all other parameters unchanged). Plotted in the same way as Fig.~\ref{fig:s1}. (J) Same as D but with larger conductance (1000 $\mu$S/c$\mbox{m}^2$) from integrators to differentiators,  so as to reimburse the higher spike threshold introduced by smaller $\beta_w$. (K) Same as F but with smaller conductance (820 $\mu$S/c$\mbox{m}^2$) from integrators to differentiators. }
    \label{fig:sweep}
\end{figure}

\begin{table}[]
    \centering
    \begin{tabular}{|c|c|}
    \hline
    Parameter               &	Value \\
    \hline
    $E_{Na}$                &	50 mV \\
    $E_K$                   &	-100 mV \\
    $E_L$                   &	-70 mV \\
    $g_{Na}$                &	20 mS/c$\mbox{m}^2$ \\
    $g_K$                   &	20 mS/c$\mbox{m}^2$ \\
    $g_L$                   &	2 mS/c$\mbox{m}^2$ \\
    $\phi_w$                &	0.15 \\
    $C$                     &	2 mF/c$\mbox{m}^2$ \\
    $\beta_w$               &	-1.2 mV \\
    $\gamma_m$              &	18 mV \\
    $\gamma_w$              &	10 mV \\
    $E_{syn}$, excitatory   &	0 mV \\
    $E_{syn}$, inhibitory   &	-90 mV \\
    \hline
    \end{tabular}
    \caption{\textbf{Parameters of the single-neuron model.}}
    \label{table:1}
\end{table}

\begin{table}[]
    \centering
    \begin{tabular}{|c|c|c|c|}

    \hline
    Parameter	    &   Heterogeneous                   &	Differentiator PN           &	Integrator LHN \\
    \hline
    $\beta_w$, ORN	&   -23 mV	                        &   -23 mV	                    &   -23 mV \\
    $\beta_w$, PN	&   5 mV	                        &   -19 mV	                    &   5 mV \\
    $\beta_w$, LHN	&   -19 mV	                        &   -19 mV	                    &   5 mV \\
    $\sigma_V$, ORN	&   38 $\mu$A/c$\mbox{m}^2$	        &   38 $\mu$A/c$\mbox{m}^2$     &	38 $\mu$A/c$\mbox{m}^2$ \\
    $\sigma_V$, PN	&   38 + 15$\eta$ $\mu$A/c$\mbox{m}^2$&   15 $\mu$A/c$\mbox{m}^2$   &	38 + 15$\eta$ $\mu$A/c$\mbox{m}^2$ \\
    $\sigma_V$, LHN	&   15 $\mu$A/c$\mbox{m}^2$	        &   15 $\mu$A/c$\mbox{m}^2$     &	38 + 15$\eta$ $\mu$A/c$\mbox{m}^2$ \\
    $g_{syn}$, PN   &   345 $\mu$S/c$\mbox{m}^2$	    &   1170 $\mu$S/c$\mbox{m}^2$   &	345 $\mu$S/c$\mbox{m}^2$ \\
    $g_{syn}$, LHN  &   975 $\mu$S/c$\mbox{m}^2$	    &   715 $\mu$S/c$\mbox{m}^2$    &	285 $\mu$S/c$\mbox{m}^2$ \\
    \hline

    \end{tabular}
    \caption{\textbf{Parameters of the \textit{Drosophila} AL network model.} $\eta$ is a random number sampled from a uniform distribution ranging from 0 to 1. }
    \label{table:2}
\end{table}

\begin{table}[]
    \centering
    \begin{tabular}{|c|c|c|c|}
    \hline
    Parameter	            &   Heterogeneous               &	Differentiator PN               &	Integrator LHN \\
    \hline
    $\beta_w$, Input        &   -23 mV                      &	-23 mV	                        &   -23 mV \\
    $\beta_w$, Even	        &   -19 mV                      &	-19 mV                          &	5 mV \\
    $\beta_w$, Odd	        &   5 mV                        &   -19 mV                          &	5 mV \\
    $\sigma_V$, Input       &	38 $\mu$A/c$\mbox{m}^2$     &	38 $\mu$A/c$\mbox{m}^2$	        &   38 $\mu$A/c$\mbox{m}^2$ \\
    $\sigma_V$, Even        &	15 $\mu$A/c$\mbox{m}^2$     &	15 $\mu$A/c$\mbox{m}^2$	        &   38 + 15$\eta$ $\mu$A/c$\mbox{m}^2$ \\
    $\sigma_V$, Odd         &	38 + 15$\eta$ $\mu$A/c$\mbox{m}^2$  &	15 $\mu$A/c$\mbox{m}^2$ &	38 + 15$\eta$ $\mu$A/c$\mbox{m}^2$ \\
    $g_{syn}$, Even         &	975 $\mu$S/c$\mbox{m}^2$	&   975 $\mu$S/c$\mbox{m}^2$        &	345 $\mu$S/c$\mbox{m}^2$ \\
    $g_{syn}$, Odd          &	345 $\mu$S/c$\mbox{m}^2$	&   975 $\mu$S/c$\mbox{m}^2$        &	345 $\mu$S/c$\mbox{m}^2$ \\
    \hline
    \end{tabular}
    \caption{\textbf{Parameters of the deep FFN model}. Even and Odd represent the $2n$ and $(2n+1)$-th layer where $n=1,2,…,5$, respectively. $\eta$ is a random number sampled from a uniform distribution ranging from 0 to 1. }
    \label{table:3}
\end{table}
\end{appendices}
\end{document}